\documentclass{PoS}
\def\scr#1{\mbox{\scriptsize #1}}

\title{What we have learned  so far from 3-fluid 
hydrodynamics}

\ShortTitle{What we have learned so far from 3-fluid 
hydrodynamics}

\author{\speaker{Yu.B. Ivanov} and V.N. Russkikh\\
        Kurchatov Institute, Moscow, Russia \\
Gesellschaft  f\"ur Schwerionenforschung,
Darmstadt, Germany \\       
E-mail: \email{Y.Ivanov@gsi.de}, \email{russ@ru.net}}

\abstract{
Available data on heavy-ion collisions at AGS and SPS energies are
analyzed using a 3-fluid dynamical model within a purely hadronic
scenario. We investigate the problems met in reproducing these data
within this scheme. In particular, we try to indicate those data 
which could point towards the occurrence of a phase transition into
the quark-gluon phase. We also discuss the success of the model in
reproducing the transverse-mass spectra of various hadrons. We argue
that the simultaneous reproduction of the inverse-slope parameters of 
all considered particles may imply that these particles belong to the
same hydrodynamic flow at the instant of their freeze-out rather than
that it signals the onset of a phase transition.  
}

\FullConference{Critical Point and Onset of Deconfinement
          4th International Workshop\\
		 July 9-13 2007\\
		 GSI Darmstadt,Germany}

\begin{document}

\section{Introduction}

The interest in nucleus-nucleus collisions at incident energy range
from AGS to SPS  
has been revived because the 
onset of the deconfinement phase transition is expected in this domain.
Moreover, a critical end point \cite{Asakawa} 
of the transition line of the QCD phase diagram may be accessible in such
reactions \cite{Fodor01,Stephanov99}. 
The above expectations motivated new projects at the proposed 
accelerator facilities FAIR at GSI \cite{FAIR,CBM} and 
NICA at JINR \cite{Dubna}.
The future SPS \cite{NA49-future} and RHIC 
\cite{RHIC-future,STAR-future,PHENIX-future} programs
are also devoted to the same problems.

In Ref. \cite{3FD} we introduced a 3-fluid 
dynamical (3FD) model which is suitable for simulating heavy-ion
collisions at high incident energies. 
Up to now the most extensive simulations \cite{3FD,3FDflow,3FDpt,3FDfrz}  
were performed 
with a purely hadronic equation of state (EoS)
\cite{gasEOS}. 
This EoS should serve as a reference for subsequent simulations with
more sophisticated scenarios including a phase transition.

In this talk we would like to investigate 
how good the available data from AGS and
SPS can be understood without involving the concept of deconfinement. 
This analysis is based on success and mainly failures of our simulations. 
In particular, we would like to indicate those data which 
seem to require a phase transition into the quark-gluon phase
for their reproduction.

\section{The 3FD Model}

A direct way to address thermodynamic properties of the matter
produced in nuclear collisions consists in application of
hydrodynamic simulations. However, finite
nuclear stopping power, observed at high incident
energies, points towards strong non-equilibrium effects which 
prevent a straight application of conventional hydrodynamics
especially at the initial stage of the reaction. The use 
of viscosity and thermal conductivity does not help to overcome this
difficulty, because by definition they are suitable for weak
non-equilibrium cases only. A possible way out is to employ a
multi-fluid approximation to the collision dynamics.

Unlike the conventional hydrodynamics, where a local instantaneous
stopping of projectile and target matter is assumed, a specific
feature of the dynamic 3-fluid description is a finite stopping
power resulting in the counter-streaming regime of leading
baryon-rich matter. The basic idea of a 3-fluid approximation to
heavy-ion collisions \cite{3FD} is that at each space-time
point $x=(t,{\bf x})$ the generally nonequilibrium 
distribution function of baryon-rich
matter, can be represented as a sum of two
distinct contributions, 
$f_{\scr{bar.}}(x,p)=f_{\scr p}(x,p)+f_{\scr t}(x,p)$,
initially associated with constituent nucleons of the projectile
(p) and target (t) nuclei. In addition, newly produced particles,
populating the mid-rapidity region, are associated with a fireball
(f) fluid. 
It is assumed that constituents within each
distribution are  locally equilibrated, both thermodynamically and
chemically. This assumption justifies  the term ``fluids''.
Therefore, the 3-fluid approximation is a minimal way to 
simulate the finite stopping power at high incident energies.

Our 3FD model \cite{3FD} is a
straightforward extension of the 2-fluid model with radiation of
direct pions \cite{MRS88} and the (2+1)-fluid model
\cite{Kat93}. We extend the above models in such a
way that production of the baryon-free fireball fluid is delayed 
due to a certain formation time, during
which the matter of the fluid propagates without interactions.

\begin{figure}[htb]
\begin{minipage}{0.46\linewidth}
\includegraphics[width=6.8cm]{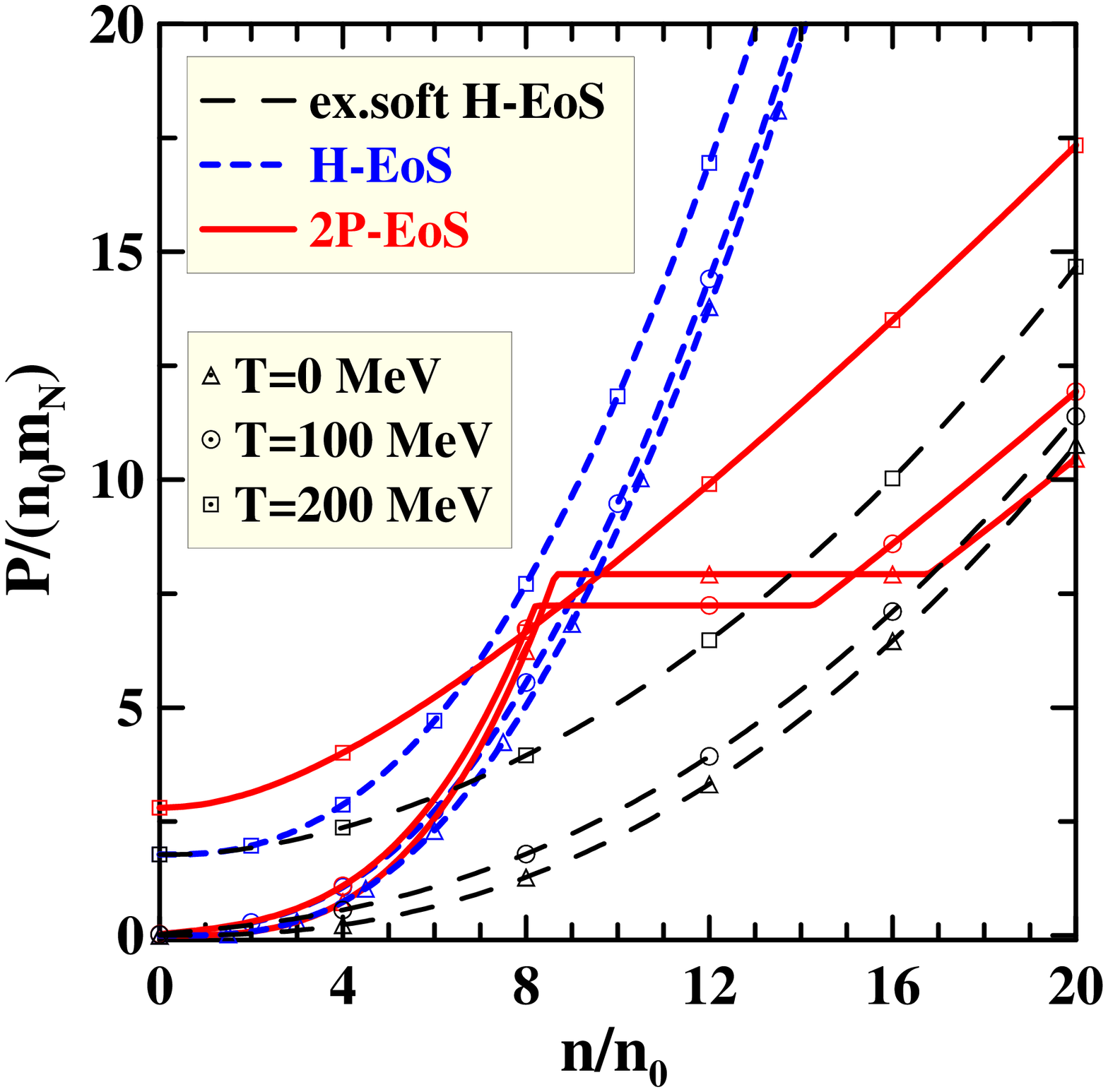}
\vspace*{-3mm}%
\caption{
Baryon-density dependence of the pressure at
  various temperatures for different EoS's \cite{gasEOS}: with  
incompressibility $K=$ 210 MeV (H-EoS) and 
$K=$ 100 MeV (ex.soft H-EoS), and 2P-EoS \cite{Toneev06}. 
Lines corresponding to different 
temperatures are tagged by different symbols, as displayed in
the figure. 
} 
\label{fig1}
\end{minipage}%
\hspace*{7mm}%
\begin{minipage}{0.46\linewidth}
\includegraphics[width=7.0cm]{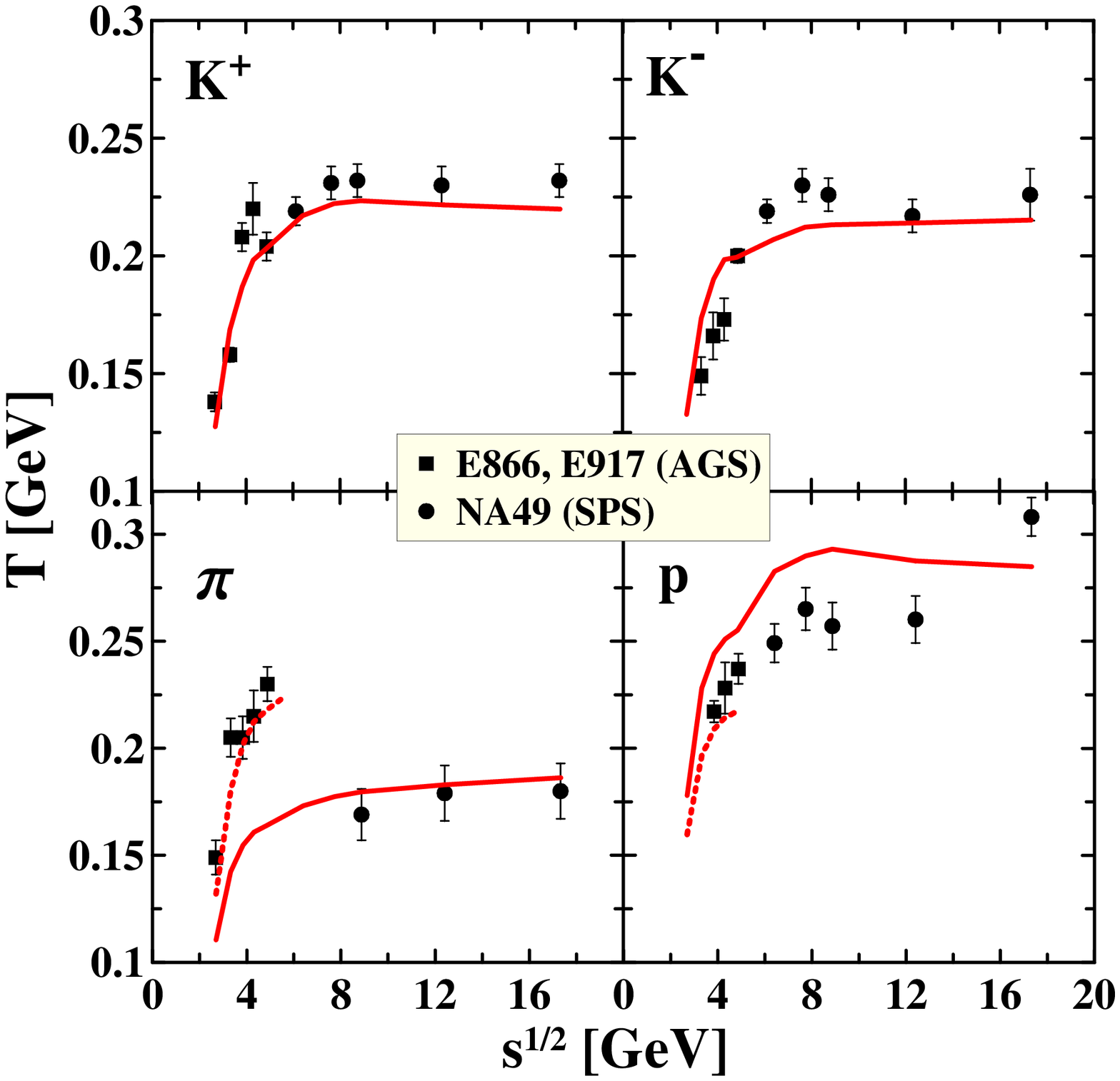}
\caption{
Inverse-slope parameters of the transverse-mass spectra of
kaons, pions and protons at midrapidity produced in central
Au+Au  and  Pb+Pb collisions as functions of
invariant incident energy. The 3FD results for  H-EoS are presented.
Solid lines correspond to the $\lambda=0$  fit, 
while dashed lines present results with $\lambda=-1$
for pions and with $\lambda=1$ for protons.
Data are from Refs. \cite{E866,NA49,E917p,NA49p}.
}
\label{fig2}
\end{minipage} 
\end{figure}

The equation of state is a key quantity for the hydrodynamic model. 
Up to now the most extensive simulations were performed 
with a purely hadronic EoS \cite{gasEOS} with the incompressibility
$K=210$ MeV (below referred as H-EoS). It does well in reproducing 
major part of the available data. The density dependence of the H-EoS
pressure at  different temperatures ($T$) is displayed in
Fig. \ref{fig1}. In this figure we  
also present an extra-soft hadronic EoS (ex.soft H-EoS) with $K=100$
MeV and a two-phase EoS (2P-EoS) \cite{Toneev06} 
which involves a 1st-order phase transition into the quark-gluon phase. 
In that 2P model, the saturation properties of symmetric nuclear matter
in the ground state are correctly reproduced. 
The  quark-gluon  phase in the 2P model is constructed as a
system of massive quasi-particles interacting via a 
density-dependent potential which simulates the hard-thermal-loop 
interactions. The 2P-EoS is in  quite reasonable agreement
with  lattice QCD data  on temperature and baryon
chemical potential dependence of relevant thermodynamic
quantities. 

Note that the phase transition leads to a softening of the EoS at 
high baryon densities. The peculiarity of the ex.soft H-EoS is that 
it is similar to the 2P-EoS in the quark-gluon phase.
In this talk we are going to discuss the reproduction of data in the
context of required softening/hardening of the EoS. 
In view of Fig. \ref{fig1}, required softening 
may indicate a phase transition into quark-gluon phase.
All other parameters of 
the model were kept fixed as described in Ref. \cite{3FD}: 
(i) the freeze-out energy density is $\varepsilon_{\scr{frz}}= 0.4$ GeV/fm$^3$, 
(ii) the friction between the baryon-rich fluids is tuned to the value
given in Ref. \cite{3FD}, 
(iii) the formation time of the fireball fluid is $\tau=$ 2 fm/c.

\section{What we have learned from analysis of data}
\label{analysis}

With the simple H-EoS we succeeded to reasonably
reproduce a great 
body of experimental data in the incident energy range
$E_{\scr{lab}}\simeq$ (1--160)$A$ GeV. The list includes 
 rapidity distributions \cite{3FD},
 transverse-mass spectra \cite{3FDpt,3FDfrz}, and 
multiplicities of various hadrons \cite{3FD}. 
However, we also found out certain problems. 
Precisely these problems we are going to analyze. 
As an exception, we are also going to discuss the success of the 3FD
model in reproduction of transverse-mass spectra, 
since the excitation functions of inverse slopes of these spectra 
were interpreted as an indication of a phase transition.

\subsection{Transverse-Mass Spectra}
\label{transverse-mass}

The 3FD results for inverse-slope parameters of transverse-mass spectra
of kaons, pions and protons  produced in central Au+Au  and  Pb+Pb
collisions are presented in Fig. \ref{fig2}. The inverse slopes $T$
were deduced by fitting the calculated spectra by the formula
\begin{eqnarray}
\label{Ttr}
\frac{d^2 N}{m_T \; d m_T \; d y} \propto
\left( m_T\right)^\lambda
\exp \left(-\frac{m_T}{T}  \right),
\end{eqnarray}
where $m_T$ and $y$ are the transverse mass and rapidity, respectively.
Though the purely exponential fit with $\lambda=0$
does not always provide the best fit of the spectra, it allows a
systematic way of comparing spectra at different incident energies.
In order to comply with experimental
fits at AGS energies (and hence with displayed experimental points),
we also present results with $\lambda=-1$ for pions
and with $\lambda=1$ for protons.

In Ref. \cite{3FDpt} it was shown that dynamical freeze-out description 
\cite{3FDfrz}, applied to the  
3FD model \cite{3FD}, naturally explains the incident energy 
behavior of inverse-slope parameters of transverse-mass spectra observed in
experiments. This freeze-out dynamics differs
from conventionally used  freeze-out schemes. It effectively brings
about a pattern similar to that of a liquid--gas phase transition.

As seen from Fig. \ref{fig2}, 
the inverse-slope parameters (with the purely exponential fit, $\lambda=0$) 
at mid rapidity reveal a "step-like" behavior. 
They increase with incident energy across the AGS energy domain 
and then saturate at SPS 
energies. In Refs. 
\cite{Gorenstein03,Mohanty03} this saturation was
associated with the deconfinement phase transition. This assumption
was indirectly confirmed by the fact that microscopic transport
models (HSD and UrQMD \cite{Bratkovskaya}, and GiBUU \cite{Wagner}), 
based on hadronic degrees of freedom, failed to reproduce
the observed behavior of the inverse slopes of kaons.
However, these transport models
\cite{Bratkovskaya,Wagner}   do 
well describe pion and proton transverse-mass 
spectra in a wide range of incident energies. 
Therefore, the failure with the kaon inverse slopes may be interpreted as
a signature  
that kaon interaction cross sections (in microscopic
models) are not big enough in order to have the kaons be captured by
matter in a common flow.  
Another possibility is that multi-body collisions are important in the
transport. This was checked within the GiBUU model \cite{Larionov07},
where three-body interactions were included in simulations.  
It was found that the three-body collisions indeed result in good
reproduction of all transverse-mass spectra \cite{Larionov07}.

Therefore, to our mind, the 
simultaneous reproduction of inverse-slopes of all considered
particles 
implies that these particles belong to 
the same hydrodynamic flow at the instant of their freeze-out.

\begin{figure}[htb]
\begin{center}
\includegraphics[width=7cm]{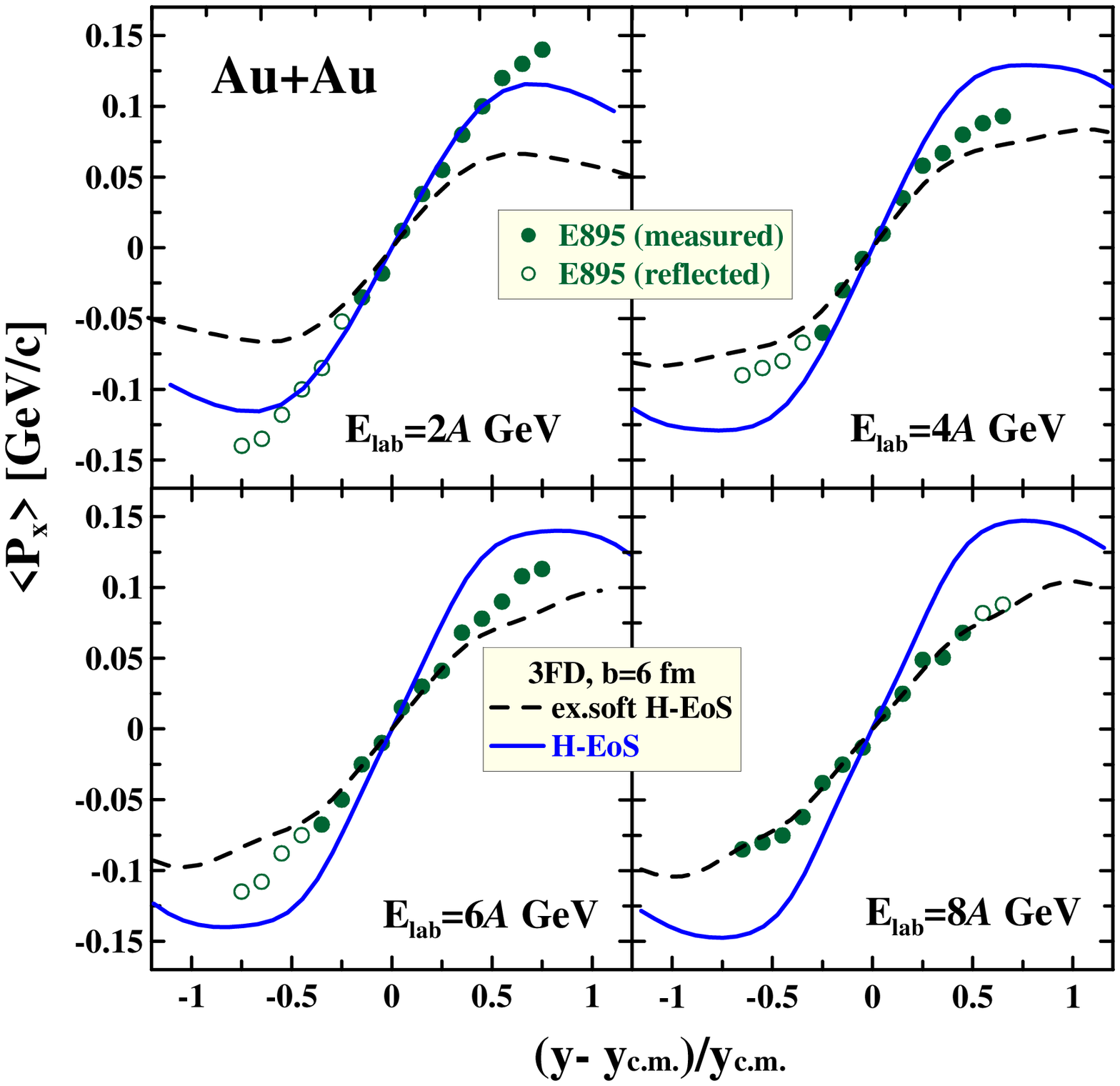}
\hspace*{7mm}%
\includegraphics[width=4.6cm]{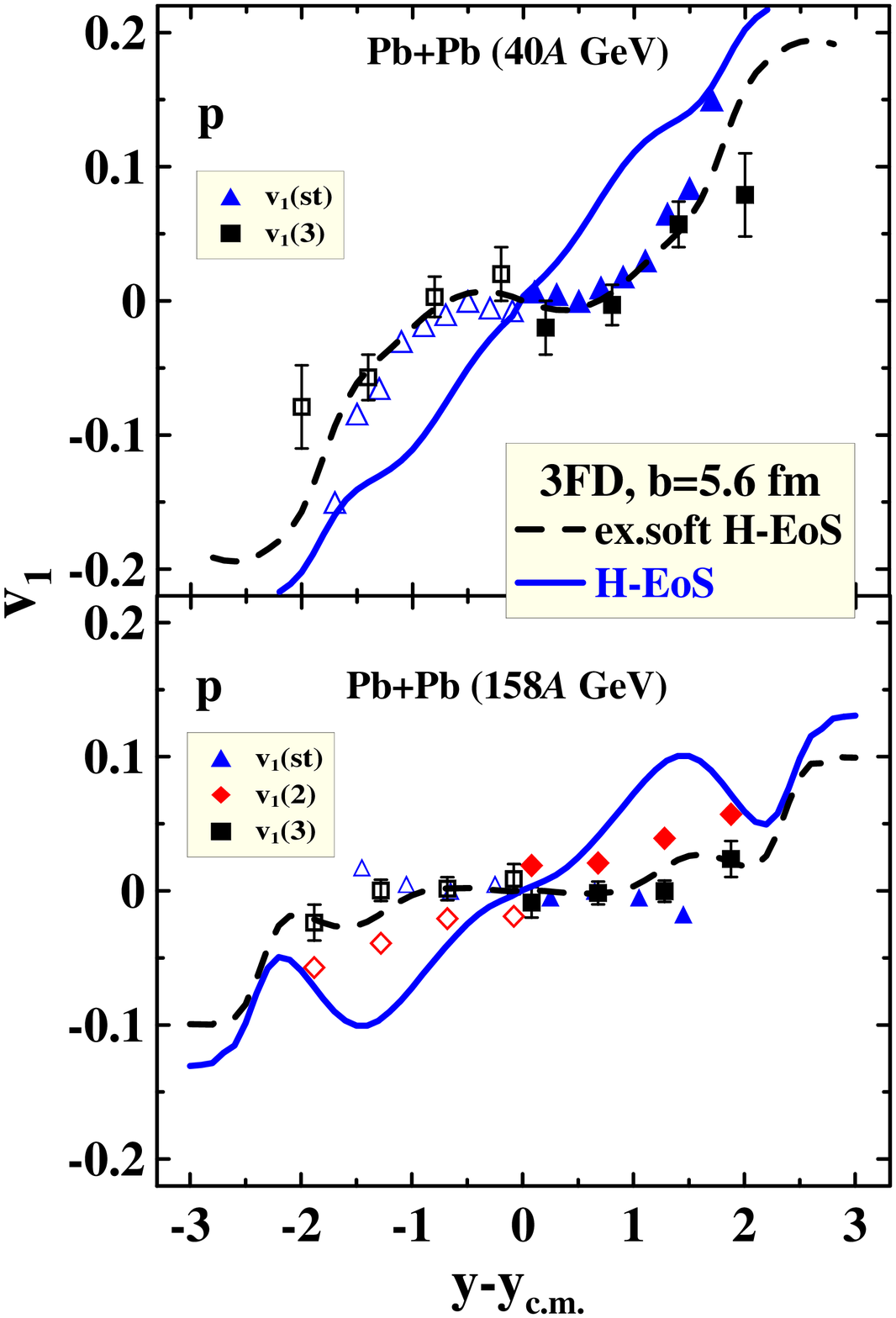}
\end{center}
\vspace*{-7mm}%
\caption{
Directed flow of protons as a function of rapidity 
for mid-central 
collisions at AGS 
(left panels) and 
SPS energies (right panels). 
The AGS data are from E895 collaboration \cite{E895-00}. 
At SPS energies, the NA49 data  \cite{NA49-03-v1} obtained by two different
methods are displayed: by the standard method ($v(st)$) 
and by the method of $n$-particle correlations ($v(n)$). 
Full symbols correspond to
measured data, while open symbols are those reflected  
with respect to the midrapidity. 
} 
\vspace*{-4mm}%
\label{fig3}
\end{figure}

\subsection{Directed Flow}
\label{Directed Flow}

Let us proceed to problems which we met.

The directed flow ($P_x$ or $v_1$) is a quantity that is
sensitive to the presence of a  phase transition. 
As 
demonstrated in Refs. 
\cite{Rischke95,Brachmann,Paech01}, the 1st-order phase transition
leads to a 
significant reduction of the directed flow
\cite{Rischke95} and even develops an antiflow behavior in the
midrapidity region \cite{Brachmann,Paech01}.

Analyzing flow data within the 3FD model \cite{3FDflow} we found that 
the directed flow data favor a steady softening of the 
EoS with increasing beam energy. 
As seen from Fig. \ref{fig3}, 
this EoS softening essentially occurs
across the AGS energy range, while at higher (SPS) energies the same extrasoft
EoS remains preferable. 
In view of above the mentioned predictions \cite{Rischke95,Brachmann,Paech01}, 
this EoS softening can be viewed as a signal of the  deconfinement
transition.

\begin{figure}[htb]
\begin{minipage}{1.0\linewidth}
\begin{center}
\includegraphics[width=12.7cm]{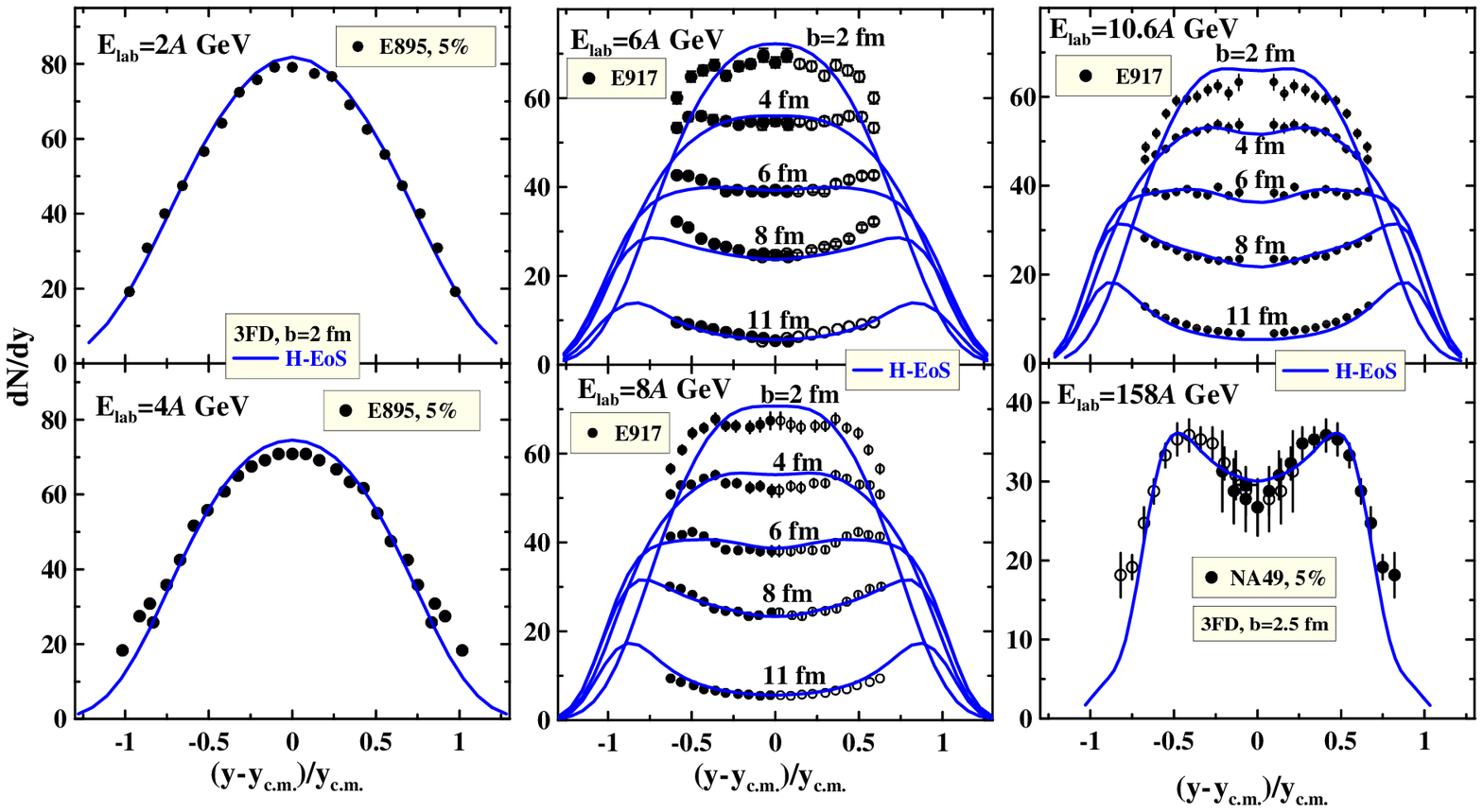}
\end{center}
$\;$\vspace*{-12mm} 
\caption{Proton rapidity spectra (solid lines) at AGS and SPS energies for 
various impact parameters ($b$) calculated with  the H-EoS. 
Experimental points are taken from 
\cite{E895:piKp} at 2$A$ and 4$A$ GeV,  
and \cite{E917:piKp} at 6$A$, 8$A$ and 10.5$A$ GeV. 
The NA49  data  for $E_{\scr{lab}}=158A$ 
GeV are from Refs.~\cite{NA49-1,NA49-04}. 
The percentage indicates 
the fraction of the total reaction cross section, 
corresponding to experimentally selected events.
} 
\label{fig5}
\end{minipage}
\\ 
%
%
\begin{minipage}{1.0\linewidth}
\vspace*{5mm} 
\begin{center}
\includegraphics[width=4.5cm]{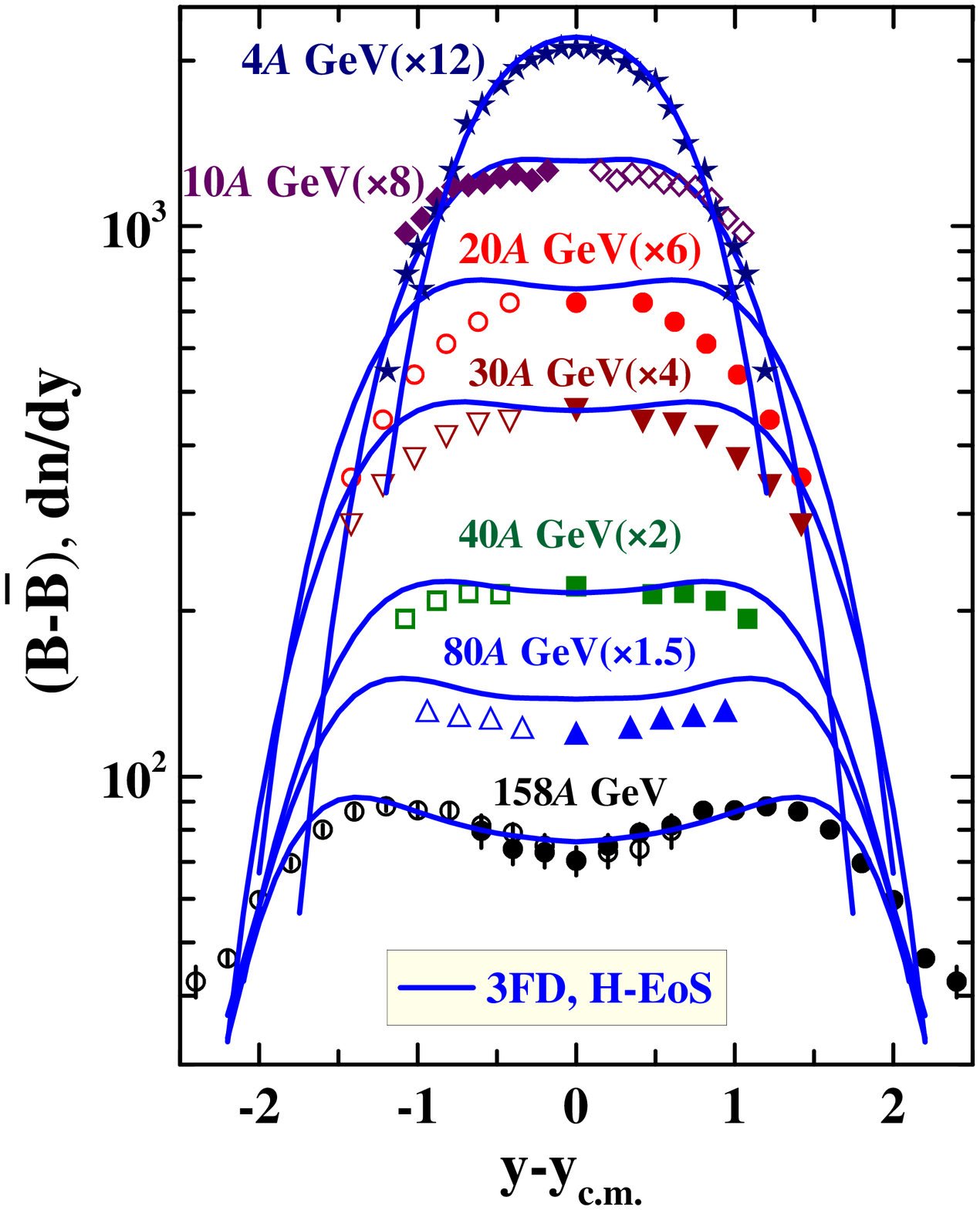}
\includegraphics[width=4.5cm]{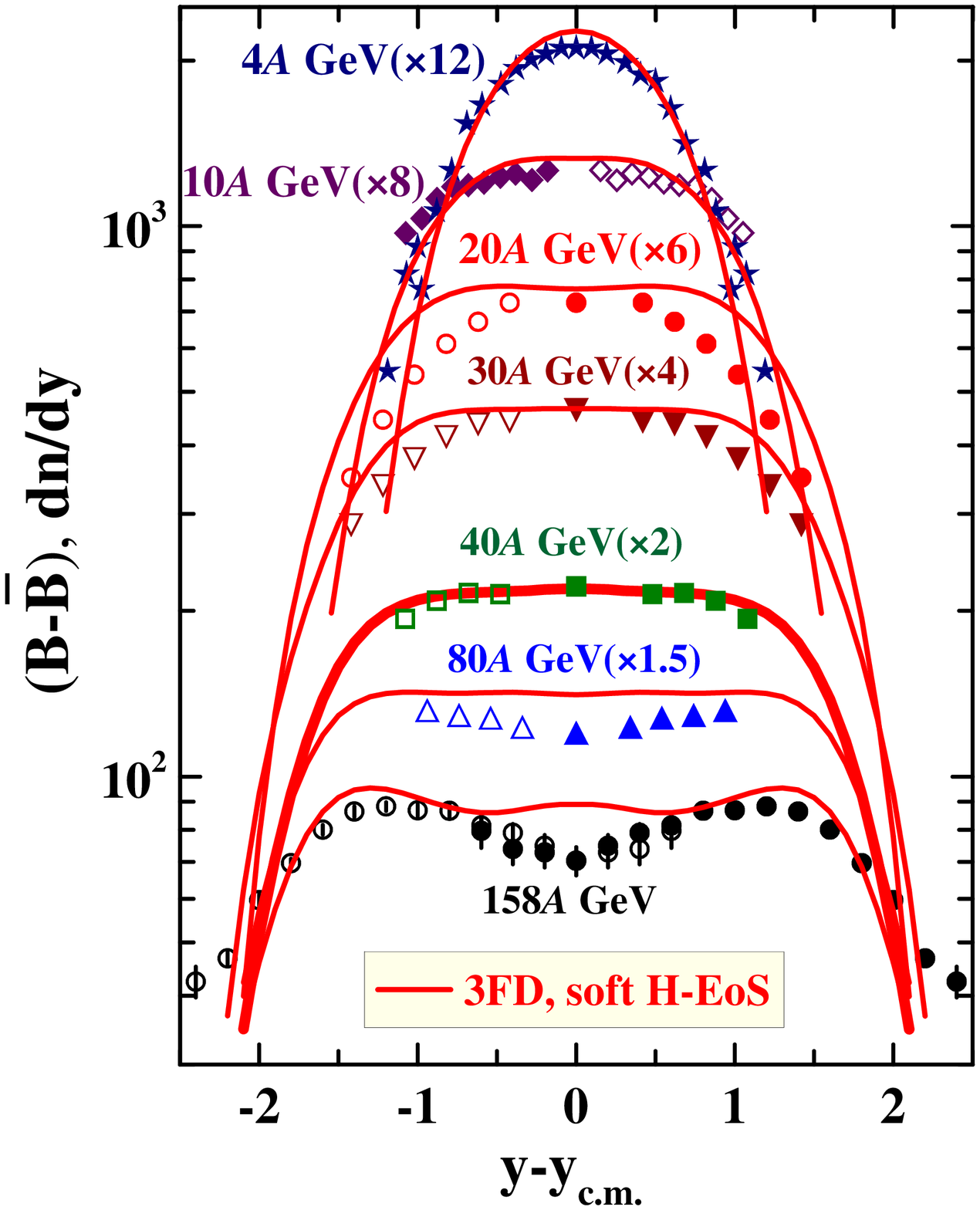}
\includegraphics[width=4.5cm]{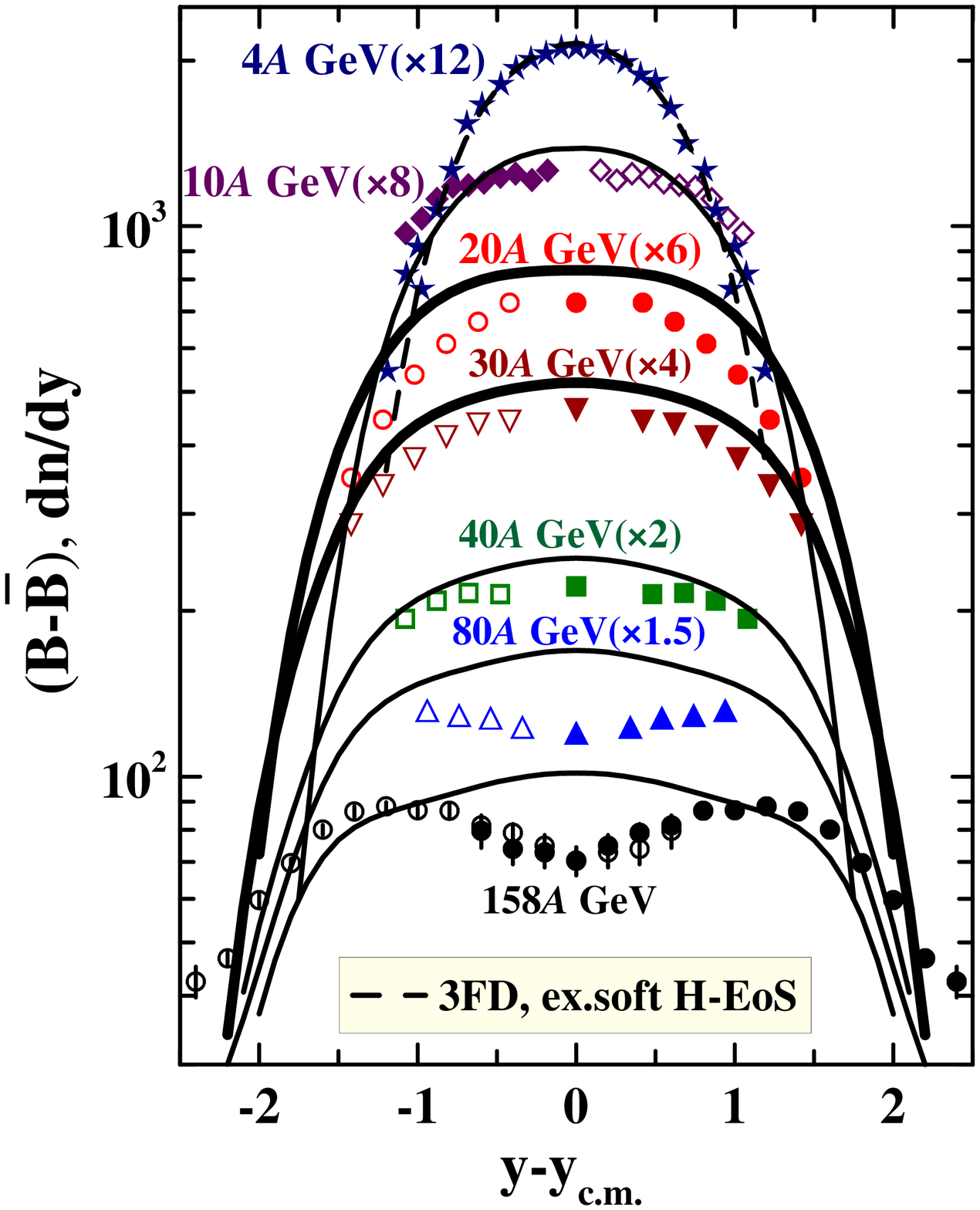}
\end{center}
$\;$\vspace*{-12mm} 
\caption{
Net-baryon rapidity spectra from  
central collisions of Au+Au 
at AGS 
energies and Pb+Pb 
at SPS energies as functions of
rapidity 
for H-EoS with $K=210$ MeV 
(left  panel), soft H-EoS with $K=150$ MeV (middle panel) and 
ex.soft H-EoS with $K=100$ MeV (right panel). 
Data at SPS energies are from Ref.  \cite{NA49-1,NA49-04,NA49-07}. 
Net-baryon data at AGS energies are obtained from the corresponding 
proton data (Ref. \cite{E895:piKp} at 4$A$ GeV and Ref.   
\cite{E917:piKp} at 10.5$A$ GeV) by multiplying them by the factor
$A/Z$. For clarity of
    representation, data sets from bottom
    to top are scales by additional factors displayed in the figure. 
\vspace*{-5mm} 
} 
\label{fig6}
\end{minipage}%
\end{figure}

\subsection{Rapidity Distributions}
\label{Rapidity}

 The proton 
rapidity distributions basically reflect the stopping power achieved in the 
nuclear collision. Therefore, their reproduction would indicate that a model 
properly describes at least the global features of the collision process. 
Till the beginning of this year we reported success of the 3FD model
with H-EoS in reproduction of the rapidity distributions, see Fig. \ref{fig5}.
However, when new data by the NA49 collaboration \cite{NA49-07} were
then published , we saw that the situation is not that evident, 
see Fig. \ref{fig6}, left panel. In spite of fitting rapidity spectra
at reference energies of 10$A$ and 158$A$ GeV, the H-EoS failed to
reproduce the net-baryon distributions  at energies in-between. 

We studied the effect of the stiffness of the EoS on these  distributions,
see Fig. \ref{fig6}. 
It turned out that the net-baryon rapidity spectrum at 40$A$ GeV
is almost perfectly reproduced with the 
soft H-EoS with the incompressibility $K=$ 150 MeV, while ex.soft
H-EoS with $K=$ 100 MeV 
is required at energies 20$A$ and 30$A$ GeV. For the energies 80$A$
and 158$A$ GeV 
the best result is still achieved with our standard H-EoS. At the same time, 
rapidity spectra at AGS energies are quite insensitive to the stiffness.

Thus, we conclude that similarly to the directed flow
the net-baryon rapidity spectra at energies 20$A$--40$A$ 
also require a softer EoS. 
This may be a signal of deconfinement transition.
However, the fact that the standard H-EoS remains favorable for
rapidity spectra  
at 80$A$ and 158$A$ GeV apparently contradicts the trend found in the 
directed-flow data. This contradiction can be an artifact of a biased fit to the data: 
we have changed only the stiffness while keeping other quantities 
(the freeze-out energy density, the friction, and the formation time) fixed. 
It would be of interest to try an unbiased fit of the SPS data 
in order to find out if these data are compatible with a soft EoS.

\begin{figure}[thb]
\begin{center}
\includegraphics[width=8.5cm]{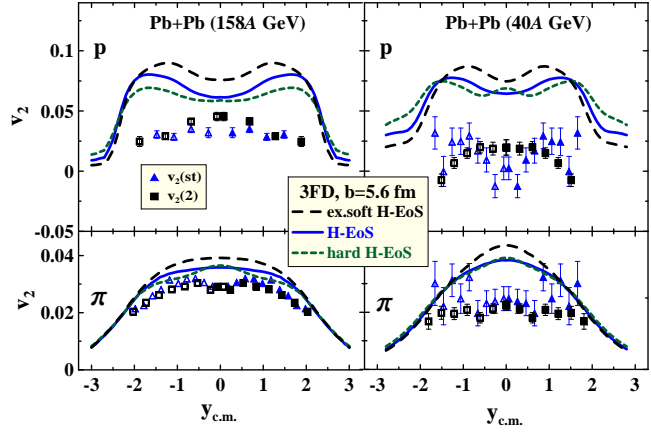}
\end{center}
\vspace*{-5mm} 
\caption{
Elliptic flow of protons (upper panels) and pions (lower panels) 
as a function of rapidity 
in mid-central ($b=$ 5.6 fm) Pb+Pb collisions. 
3FD calculations with H-EoS, ex.soft H-EoS, and hard H-EoS ($K=380$ MeV) are presented. 
The NA49 data \cite{NA49-03-v1} obtained by two different
methods are displayed: by the standard method ($v(st)$) 
and by the method of $n$-particle correlations ($v(n)$). 
Full symbols correspond to
measured data, while open symbols are those reflected  
with respect to the midrapidity. 
}
\vspace*{-3mm} 
\label{fig7}
\end{figure}

\subsection{Elliptic Flow}
\label{Elliptic Flow}

In Ref. \cite{3FDflow} we  found that it is impossible to simultaneously
reproduce the directed  and elliptic flow with the same EoS. 
The directed flow requires a softer EoS,
while the elliptic flow demands for a harder one. 
In that paper we assigned this deficiency to a 
lack of the proper description of the nonequilibrium
transverse-momentum anisotropy at the initial 
stage of nuclear collision. 

Later we became aware that the elliptic flow is strongly 
affected by a post-hydro cascade (so called ``afterburner'') 
\cite{Teaney01,Hirano07}. 
This afterburner essentially reduces (approximately twice at SPS energies) 
the $v_2$ values
achieved during the hydrodynamic stage. In particular, this reduction
may help to bring the 3FD results in correspondence with the
experimental data  at 158$A$ GeV. 
In Fig. \ref{fig7} we present purely hydrodynamic predictions
for the elliptic flow based on the H-EoS's with different stiffnesses, 
keeping in mind that they will be approximately reduced 
to half the value by the  post-hydro cascade. 
However, the model still strongly overestimates the proton $v_2$ at 40$A$ GeV
(even on account of the afterburner effect). 

Considering that $v_2$ at 40$A$ and 158$A$ GeV only weakly depends 
on the stiffness, see Fig. \ref{fig7}, we cannot associate this
``collapse'' of the proton $v_2$  
at 40$A$ GeV with an onset of the phase transition. 
In Ref. \cite{Shuryak05} this ``collapse''  was assumed to be a signal 
of the critical point in the quark-gluon phase diagram.

\begin{figure}[htb]
\begin{center}
\includegraphics[width=11.5cm]{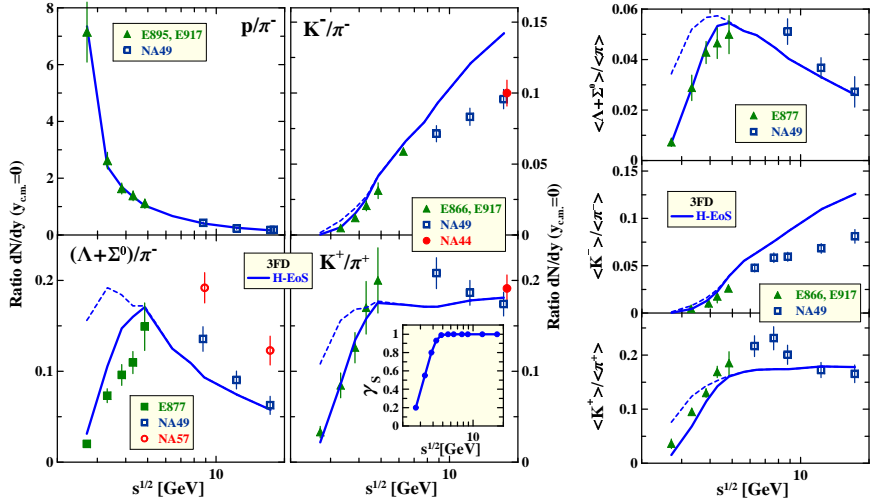}
\end{center}
\vspace*{-5mm} 
\caption{
Excitation functions of ratios of midrapidity yeilds (left block of panels) and 
total yields (right block of panels) of hadrons produced in 
central collisions of Au+Au at AGS energies and Pb+Pb at SPS
energies. 3FD results are presented for H-EoS. 
Bold lines correspond to calculations with additional strangeness
suppression $\gamma_S$ (see insertion in the left  block of panels), 
while thin lines --- to those without suppression. 
}
\vspace*{-4mm} 
\label{fig8}
\end{figure}

\subsection{Hadron Ratios}
\label{Hadron Ratios}

The hadron ratios of hadron abundances arouse interest because 
the observed maximum in the $K^+/\pi^+$ ratio can be interpreted as a signal of
the onset of the phase transition into quark-gluon phase
\cite{gaz,gaz2,bra,vk2}. In 
addition, these ratios most distinctly demonstrate discrepancies in
description of hadron multiplicities. Fig. \ref{fig8}
presents hadron ratios in the midrapidity region (i.e. ratios of
midrapidity densities of particles) and those
of total multiplicities calculated within
the 3FD model with H-EoS, as well as their comparison with experimental data.

It is immediately seen that the strangeness production at low incident
energies is evidently overestimated within the 3FD model. This is not
surprising, since the hadronic-gas EoS used at the freeze-out
\cite{3FD} is based on the grand canonical ensemble. This shortcoming
can be easily curied by introduction of a phenomenological factor
$\gamma_S$ (see, e.g., Refs. \cite{Becattini05}), which
accounts for an additional strangeness 
suppression due to constraints of canonical ensemble. 
The resulting $\gamma_S$ factor is presented in the inserted panel of 
Fig. \ref{fig8}. As seen, at $E_{\scr{lab}}>$ 10$A$ GeV there is no
need for additional strangeness suppression. 

The 3FD model with H-EoS reproduces hadron
ratios approximately to the same extent 
as transport models HSD and UrQMD \cite{bra} and GiBUU
\cite{Larionov07} based on  
hadronic degrees of freedom do. In particular, the 3FD model also
fails to reproduce the 
maximum in the $K^+/\pi^+$ ratio. We found that  calculated 
 hadron ratios are quite insensitive to stiffness of the EoS.

 Therefore, 
the observed maximum in the $K^+/\pi^+$ ratio can hardly be a signal
of the 
onset of a phase transition. This may be a signal of the critical point
in the phase diagram. Indeed, relaxation processes are expected 
to be essentially slowed down near the critical point. If it concerns also 
chemical equilibration, the resulting hadron ratios would reflect the
chemical content 
of the system at earlier stages of its evolution. Then the $K^+/\pi^+$ ratio 
would turn out to be enhanced near the critical point.

\section{Summury}
\label{Summury}

In these work we analyzed 
how good the available data from AGS and
SPS can be understood within a purely hadronic scenario. 
The analysis is based on failures of our simulations
of heavy-ion collisions within the 3FD model with a purely hadronic EoS. 
We tried to cure the problems in the reproduction of various
observables by means  
of changing the stiffness of the hadronic EoS. Note that in terms of
hydrodynamics, a   
softening of the EoS can be a manifestation of a phase transition. 
Our results are as follows: 
\begin{itemize}
\item
Directed flow data (at top AGS and SPS energies) 
and net-baryon rapidity spectra (at low SPS energies) favor
a softer EoS. 
This required softening may be interpreted as an indication 
of the deconfinement transition. 
\item
The ``Collapse'' of the proton elliptic flow at 40$A$ GeV and 
observed maximum in the $K^+/\pi^+$ ratio hardly signal an 
onset of a phase transition, since these quantities are
quite insensitive to the stiffness of the EoS. 
This observations  may signal a proximity of a critical point
in the phase diagram. 
\end{itemize}

In this work we discussed problems of the hadronic scenario but with
one exception.  
This exception is the success of the 3FD model in reproduction 
of transverse-mass spectra of various hadrons (in particular, kaons), 
since the excitation functions of inverse slopes of kaon spectra 
were interpreted as an indication of a phase transition
\cite{Gorenstein03,Mohanty03}.  Our conclusion is as follows: 
\begin{itemize}
\item
The simultaneous reproduction of the inverse-slopes of all considered
particles ($p$, $\pi$ and $K$) within the 3FD model 
suggests that these particles belong to 
the same hydrodynamic flow at the instant of their freeze-out 
rather than signals an onset of a phase transition. 
\end{itemize}

3FD simulations with 2P-EoS \cite{Toneev06}, 
which involves the 1st-order phase transition into quark-gluon phase, are now 
in progress.

We are grateful to J. Knoll for stimulating discussions and for
critical reading the text of this paper. 
This work was supported in part by the Deutsche  
Forschungsgemeinschaft (DFG project 436 RUS 113/558/0-3), the
Russian Foundation for Basic Research (RFBR grant 06-02-04001 NNIO\_a),
Russian Federal Agency for Science and Innovations 
(grant NSh-8756.2006.2).

\end{document}